# The investigation of social networks based on multi-component random graphs

V. N. Zadorozhnyi
Omsk state technical university
E-mail: zwn@yandex.ru

E. B. Yudin
Sobolev Institute of Mathematics, Siberian Branch of the Russian Academy of Sciences
Novosibirsk, Russia
E-mail: udinev@asoiu.com

*Abstract*—**The methods of non-homogeneous random graphs calibration are developed for social networks simulation. The graphs are calibrated by the degree distributions of the vertices and the edges. The mathematical foundation of the methods is formed by the theory of random graphs with the nonlinear preferential attachment rule and the theory of Erdős–Rényi random graphs. In fact, well-calibrated network graph models and computer experiments with these models would help developers (owners) of the networks to predict their development correctly and to choose effective strategies for controlling network projects.**

## I. Introduction

For the last two decades, the mathematical foundation of Network Science – random graphs theory – has been enriched by a developed mathematical apparatus providing an opportunity to construct and investigate various graph models of real growing networks. The theory [1, 2] of random graphs with the nonlinear preferential attachment rule (the NPA rule) is notable for wide opportunities of the graph models calibration by statistical data on the simulated networks. Thus, the NPA graph theory allows one to form adequate models for a wide class of networks and apply these models to predict the course of the studied networks development and/or to design effective strategies for the network project control.

The NPA graph is grown from a small 'seed'-graph by adding another graph increment – a new vertex with random number $x$ of outgoing arcs at moments $t = t_1, t_2, \dots$ . The ends of the graph incrrement arcs attach to any randomly selected graph vertices. Probability $p_i$ of the arc attachment to vertex $i$is in proportion with the vertex weight $f$, determined by its degree $k_i$:

$$p_i = \frac{f(k_i)}{\Sigma_j f(k_j)}, \quad i, j = 1, \dots, N, \tag{1}$$

where$N$ is the number of the graph vertices.

At infinite addition of the graph increments, an infinite graph is formed. Therefore, the NPA random graph is set by two parameters – probabilities distribution $\{r_k\} = \{r_g, \dots, r_h\}$ for a random variable $x$ (the number of the increment arcs) and weight function $f(k) \geq 0$. Moreover, $f(k) > 0$ then and only then, when $g \leq k \leq M$ (where $g \geq 0$, $M \leq \infty$). Function $f(k)$ can be specified as subscripted variable $f_k$ or as weighting sequence $\{f_k\}$.

The article reveals non-uniformity of social networks structure, and the methods accounting for this non-uniformity are developed for calibrating the NPA graphs: the graphs are calibrated simultaneously by degree distributions of vertices and edges.

## II. Theory

**Vertex degree distribution.** At the given distribution $\{r_k\}$ of the graph arcs number $x$ and the given weights $\{f_k\}$, stationary (at $t \to \infty$) graph vertex degree distribution (VDD)is determined by the recurrence formula found in [1]:

$$Q_k = \frac{r_k \langle f \rangle + m f_{k-1} Q_{k-1}}{\langle f \rangle + m f_k}, \quad k = g, g+1, g+2, \dots \tag{2}$$

where$Q_k$ is probability of a randomly selected vertex of an infinite graph having degree$k$;$m = \Sigma_k (k r_k)$ is the mean number of the arcs in the increment, and value $\langle f \rangle$ of the vertex mean weight is calculated by solving equation system (2) with additional equation

$$\langle f \rangle = \sum_{k=g}^{M} f_k Q_k . \tag{3}$$

The equation resulting from the graph growth is used as a control one. This is an equation of medium degree $\langle k \rangle$ of the doubled mean increment degree:

$$\langle k \rangle = \sum_{k=g}^{M+1} k Q_k = 2m . \tag{4}$$

A simple procedure for a quick numerical solution of

The reported is funded by RFBR as a part of the research project No. 16-31-60023 mol_a_dk system (3), (4) in spreadsheet tables (for example, in Excel) is described in [1].

**Arcs/edges degree distribution.** The problem of acrs/edges degree distribution for the NPA graphs was solved in general terms for the first time in [3]. At given $\{r_k\}$ and $\{f_k\}$, the graph is grown with stationary distribution of arc probabilities, determined by the recurrence formula

$$Q_{l,k} = \frac{f_{k-1}(lr_l Q_{k-1} + m^2 Q_{l,k-1}) + f_{l-1} m^2 Q_{l-1,k}}{m(\langle f \rangle + mf_k + mf_l)}\,,$$

$$l\,,\, k = g, g+1, g+2, \ldots, \qquad (5)$$

where $Q_{l,k}$ is the probability of a randomly selected arc outgoing the vertex with degree $l$ and ingoing the vertex with degree $k$. Matrix $\mathbf{Q} = \| Q_{l,k} \|$ is calculated by formula (5) line by line. Probabilities $Q_k$ are determined above.

If the NPA graph is used for simulating networks with undirected links, the arcs are substituted with edges. Denoting by $\Theta_{l,k}$ a stationary probability of a randomly selected end of a randomly selected edge being incident to the vertex with degree $l$, and the other end of the edge to the vertex with degree $k$, we find $\Theta_{l,k} = (Q_{l,k} + Q_{k,l})/2$. Therefore, matrix $\mathbf{\Theta} = \|\Theta_{l,k}\|$ of the graph edges degree distribution (EDD) can be calculated by matrix $\mathbf{Q} = \| Q_{l,k} \|$ of arcs degree distribution through transformation:

$$\mathbf{\Theta} = \frac{1}{2}(\mathbf{Q} + \mathbf{Q}^{\mathrm{T}})\,, \qquad (6)$$

where T is the matrix transposition symbol.

**The problem of complex calibration for the NPA graphs.** Problem statement: Given the specified VDD $\{Q_k\}$ of the simulated network (possibly smoothed one) and the known empirical EDD $\tilde{\mathbf{\Theta}} = \| \tilde{\Theta}_{l,k} \|$, one needs to find probabilities $\{r_k\}$ and weights $\{f_k\}$, inducing the graph with the assigned VDD $\{Q_k\}$ and causing minimal difference in the graph EDD $\mathbf{\Theta}$ from the given degree distribution $\tilde{\mathbf{\Theta}}$ :

$$r(\mathbf{\Theta}, \hat{\mathbf{\Theta}}) = \left( \sum_{l,k=g}^{u} (\Theta_{l,k} - \hat{\Theta}_{l,k})^2 \right)^{1/2} \rightarrow \min. \qquad (7)$$

Parameter $u > g$ is selected by the EDD diagram for the simulated network so that two-dimensional interval $[g, u]^2$ covers the area of typical values $\Theta_{l,k}$. Varied parameters in problem (7) are probabilities $\{r_k\} = r_g, \ldots, r_h$ and weighting sequence $\{f_k\}$. Unlike the solution proposed for this problem in [4], here it is unnecessary to use weights $\{f_k\}$ providing accurate realization of the given VDD $\{Q_k\}$. If the required degree distributions $\{Q_k\}$ and $\tilde{\mathbf{\Theta}}$ can be realized approximately at natural weights $f_k = k$, such weights are considered as the problem solution. And if in this case one may find an appropriate simple definition of probabilities $\{r_k\}$, such $\{r_k\}$ are preferred to optimal ones in terms of problem (7).

## III. Characteristic Features of Social Networks Structure

Publication [4] demonstrates complex calibration of a growing random graph, simulating autonomous systems (AS) network, including 22,963 nodes and 48,436 links between them. The quality of the complex graph calibration carried out due to the data on AS network is shown by the diagrams in Fig. 1. The horizontal axes in the EDD diagrams include values $l$ and $k$, while the vertical axes have probabilities $\Theta_{l,k}$ of a randomly selected graph edge being incident to the vertices with degrees $l$ and $k$ ($\Theta_{l,k} = \Theta_{k,l}$ at any $l$, $k$). When speaking of the simulated network, instead of the words *vertex* and *edge* one uses the words *node* and *link*. The comparison of the last two diagrams in Fig. 1 shows that in AS network simulation the algorithm of a complex calibration proposed in [4] led to a high-quality result. This confirms that the mechanism of AS network growth is well-described by rule (1).

When simulating *social networks*, however, it becomes obvious that applying this algorithm of the complex calibration does not lead to as good results as in engineering AS network simulation.

Let us examine, for instance, the result (Fig. 2) of applying the algorithm when simulating social network Brightkite by the data [5] on its topology. At the time of the data collection, Brightkite network included 58,228 nodes and 214,078 links.

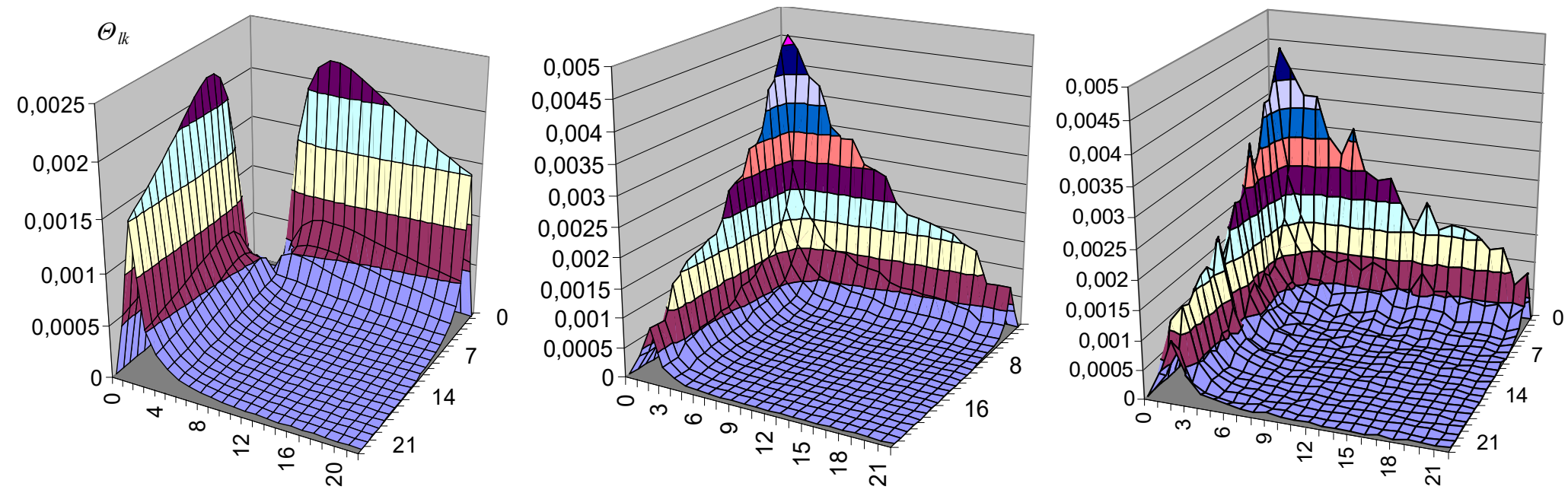

Fig. 1. The edge degree distribution of the NPA random graph (after the calibration by the vertex degree distribution – on the left, and after the complex calibration – in the center) and the degree distribution of the links between AS network nodes (on the right).

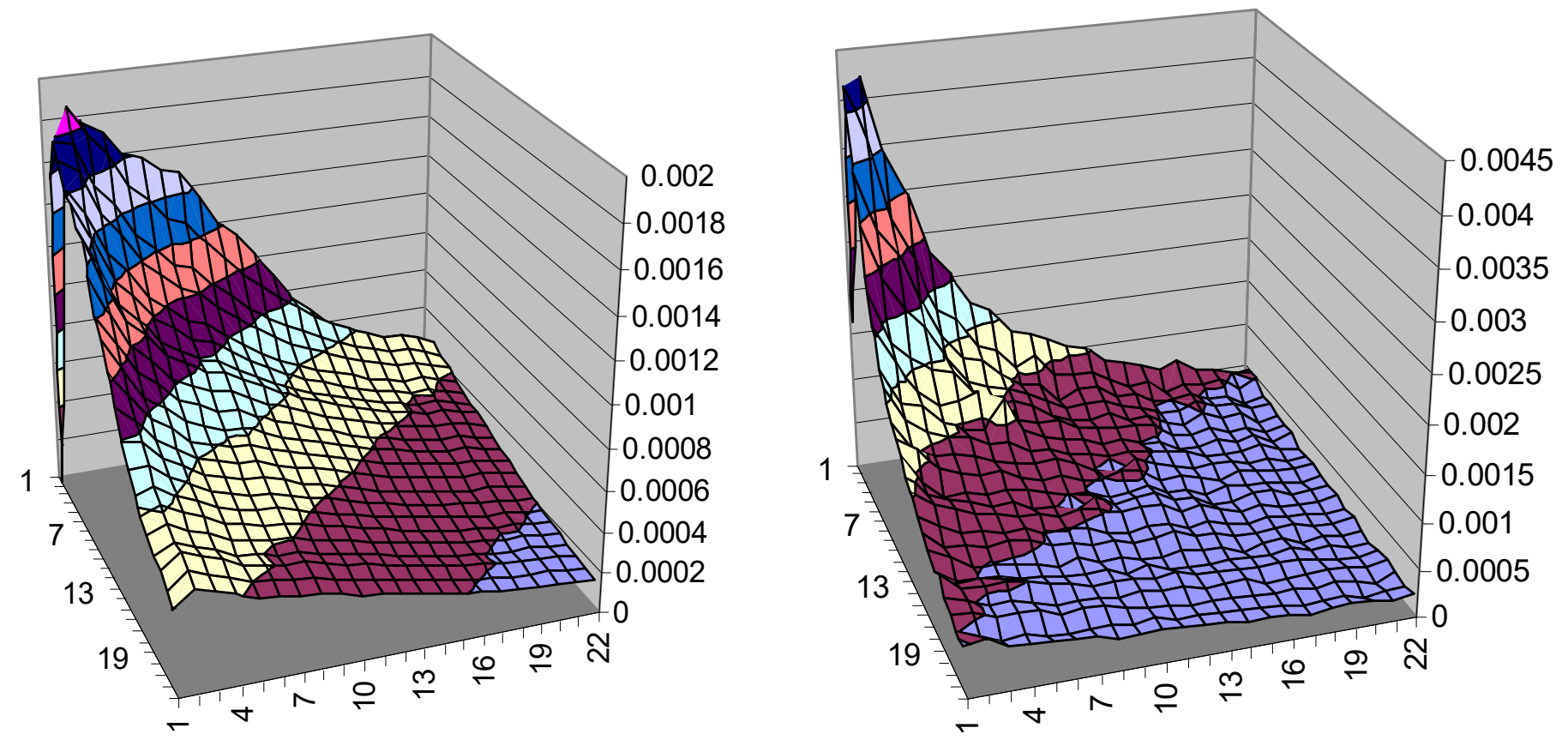

Fig. 2. The edge degree distribution for the graph calibrated by the methods [4] (on the left) and links degree distribution in Brightkite network

The dissimilarity in two diagrams' forms in Fig. 2 can still be called tolerable. But the range of probabilities $\Theta_{l,k}$ (plotted vertically) for the graph EDD is twice as narrow as that of link degree distribution for Brightkite network. Therefore, the resulting calibration cannot be considered appropriate.

As a result of the further investigation, a hypothesis arose that Brightkite network is composed of two networks with differing structures and therefore it should be simulated by composing (combining) two NPA graphs (two components), isolated or loosely connected ones. Obviously, such components can be called the components with an autonomous structure. Various suppositions on two main components of Brightkite network were studied. Significant resemblance of the graph EDD and network link degree distribution is achieved when combining the following two components. The first component is fixed, it is the Barabasi-Albert tree representing a special case of the NPA graph, specified by parameters $m = g = 1$ (i.e. $r_1 = 1$) and $f_k = k$ ($k = 1, 2, \ldots$). The second component is a complement NPA graph, with its composition minimizing the target function (7). Let the composition include $N$ vertices, and the included tree and the complement graph have $N_1 = \rho N$ and $N_2 = (1 - \rho)N$ vertices, correspondingly ($0 < \rho < 1$). It is easy to notice that VDD $\{Q_k\}$ of the composition graph is a combination of the components' VDD:

$$Q_k = \rho Q'_k + (1-\rho)Q''_k, \qquad k = 1, 2, \ldots, \tag{8}$$

where $Q'_k, Q''_k$ are the probabilities that the vertices of the first and second components have degree $k$, correspondingly. In (8) probabilities $Q'_k$ for the tree are estimated by formula (3) calculation, and probabilities $Q_k$ are given (it is a smoothed vertex degree distribution for Brightkite network). Therefore, at known $\rho$, formula (8) determines univalently probabilities $Q''_k$ of degrees $k$ for the complement graph vertices. Consequently, parameters $m$, $m'$ and $m''$ of these graphs are connected by the formula $m = \rho m' + (1 - \rho)m''$, where $m = 3.6765$ (as in a real network), $m' = 1$ (for the BA tree). This implies that at given $\rho$ we get $m'' = (m - \rho m')/(1 - \rho) = (m - \rho)/(1 - \rho)$.

In a similar way, EDD $\{\Theta_{l,k}\}$ for the two-component graph is a combination of its components' EDD:

$$\Theta_{l,k} = \gamma\Theta'_{l,k} + (1-\gamma)\Theta''_{l,k}, \qquad l,k = 1, 2, \ldots, \tag{9}$$

where $\gamma$ is the share of the tree edges in the number of the whole graph edges:

$$\gamma = m'\rho N / mN = m'\rho / m = \rho / m. \tag{10}$$

At specified $\rho$, problem (7) is solved by varying the parameters only of the complement graph. After that $\rho$ is refined, and problem (7) is solved again. In two such iterations, value $\rho$ is determined as equal to about 0.225. The resulting distribution $\{r_k\}$ of the second component includes probabilities $\{r_1, \ldots, r_{40}\}$, providing the minimum of the target function (7). The weight function of the second component is linear: $f_k = k$ ($k = 1, 2, \ldots$); this makes the model more natural as a whole. Fig. 3 shows the diagram for EDD of the graph combining two described components with an autonomous structure.

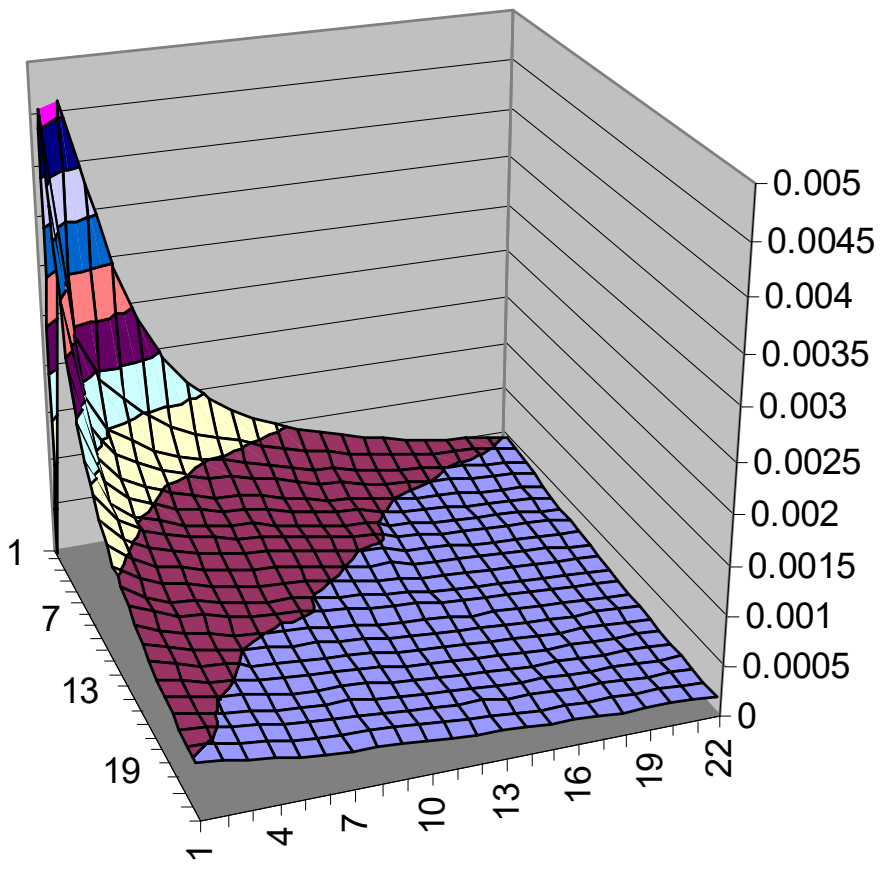

Fig. 3. The calculated edge degree distribution of the calibrated two-component graph

The EDD diagram has the form almost similar to that of the diagram for the links degree distribution (see Fig. 2 on the right) in Brightkite network (the diagram has "noises" caused by statistical errors). More accurate range coincidence of the compared diagrams can be easily achieved by a small decrease of parameter ρ, also increasing the resemblance of the diagram forms.

In all considered examples, VDD for the calibrated graphs coincides with the nodes degree distribution for the simulated networks, and the calculations validity is verified by the graph simulation modeling (SM).

The leap in the graph model accuracy as a result of combining two differing NPA graphs suggests that simulated network Brightkite is also a composition of networks with autonomous structures.

Further, the article deals with geosocial ("geocontact', "geolocational") network Gowalla. During the complex graph calibration to simulate this network, one reveals a number of characteristic structural features that cannot be explained when considering the structure as being uniform.

## IV. The hypothesis for non-uniformity of Gowalla

The data on the nodes of Gowalla network and the links between them, gathered in February, 2009 - October, 2010, are published in [6]. These data were used to calculate the degree distribution $\{Q_k\}$ of the network nodes and the degree distribution $\widetilde{\mathbf{\Theta}}$ of the links between them (Fig. 4). This figure (on the right) also displays EDD of the NPA graph, calibrated by the method proposed in [4], that provides the accurate realization of the specified VDD and uses non-linear weights. This is the best result of the complex calibration achieved by applying the method. Just like in case of Brightkite network simulation, the approach to Gowalla network as to a uniform structure led to no satisfactory calibration. Consequently, supposition arises that Gowalla network is a composition of networks with autonomous structures, as well.

## V. Unexpected problems and unexpected solutions

The attempts to obtain the graph EDD corresponding to the links degree distribution in Gowalla network by combiningtwo NPA graphs with an autonomous structure (by analogy with Brighkite network simulation) led to no success. The result of such calibration is always the EDD diagrams, having flat or concave slopes of the diagram surface and/or inappropriate ranges of values $\Theta_{l,k}$ on the vertical axis (as in the last diagram in Fig. 4). The problem (*the first one*) proved to be rather unexpected.

The convex slope of the diagram surface for the graph EDD (see the first two images in Fig. 4) can result if one of the graph components is theErdős–Rényi graph (the ER graph) [7], i.e. "classical" random graph. But the component representing the NPA graph with linear weights $f_k$ must be present as well. It ensures necessary slow (as a power function) decrease in probabilities $Q_{l,k}$ with increase in $l$ and $k$, observed in Gowalla network.

The other two problems arose when searching for such components (the ER graph and the NPA graph with linear weights), their composition would allow one to obtain EDD close to the links degree distribution in Gowalla network. One of these problems is as follows: even though the ER component allows one to achieve rather good resemblance of the graph forms for the compared degree distributions, it makes a too steep slope of the diagram surface for EDD (much steeper than necessary). Moreover, it cannot be compensated by selecting an appropriate second component (the NPA graph). Another problem is that it is impossible to obtain a sharp peak on the diagram surface for EDD (see the last diagram in Fig. 4) when calibrating.

When comparing Poisson degree distribution of the ER graph vertices with the standard one, it is obvious that one needs to increase VDD dispersion to obtain more gentle slope on the surface of EDDdiagram. It should be made, however, without changing the mathematical expectation of the degree distribution. But it is impossible since the mathematical expectation of Poisson distribution equals the dispersion. To solve this problem, the concept "autocorrelated ER graph" (the AER graph) was adopted. When constructing this graph between random attachments of vertex pairs, positive correlation was introduced. Using the AER graph as a component solved all three problems: the surface of the EDD diagram was convex one at calibration, its slope became more gentle, and the surface peak turned out sharp.

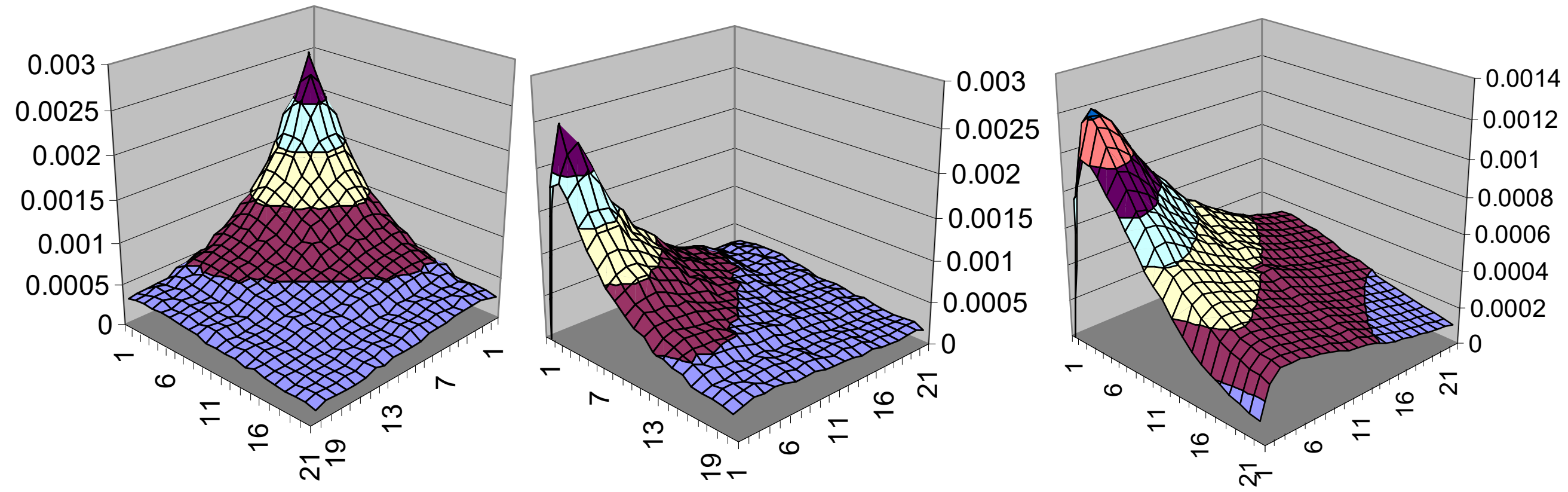

Fig. 4. The links degree distribution in Gowalla network is on the left and in the center (two perspectives of the three-dimentional diagram), the best result of the uniform NPA graph calibration is on the right

## VI. THE DESCRIPTION OF THE CALIBRATED GRAPH

The graph on $N$ vertices, calibrated by the data on Gowalla network consists of two components. The first component is the AER graph on $N_1 = \rho N$ vertices, where $\rho \approx 0.35$. The second component is the NPA graph on $N_2 = (1 - \rho)N$ vertices. If taking $N$ =100 000, we get $N_1$ = 35000, $N_2$ = 65 000.

The AER component is constructed in the following way. $N_1$ isolated vertices are taken. The edge is lead from vertex $i = 1$ to vertex $j = i + 1$ (then to vertex $j = i + 2$, $j = i + 3$, …, $j = N_1$) with probability $p = (p_a + z_{j-1})/2$, where $p_a = a/(N_1 - 1)$; $a = 2.75$ is the mean vertex degree. Random variable $z_{j-1} = 0$, if the edge is not traced to vertex $j - 1$, and $z_{j-1} = 1$, if the edge is led to vertex $j - 1$. Then, the edges connecting vertex $i = 2$ (vertex $i = 3, 4, …, N_1 - 1$) to vertices $j = i + 1, i + 2, …, N_1$ are distributed in a similar way. After the graph having been grown, all isolated vertices are deleted from it, as well as isolated vertex pairs connected by a single edge.

The second component, the NPA graph, is grown at linear weights $f_k = k$ and simple limited degree distribution $\{r_k\}$, determined by formula $r_k \approx 0.3004(k - 0.1259)^{-1.2562}$, $k = 1, …, 50$.

The diagram of the calculated EDD for the described two-component graph is shown in Fig. 5 on the left (deleting isolated pairs of the connected vertices was not taken into account in calculation). In the center there is EDD for the graph, grown by simulation modeling. Comparing these diagrams of the graph EDD with the diagram of the links degree distribution inGowalla network (see Fig. 4 in the center) and considering the manual selection of the components parameters for the calibrated graph, the conclusion can be made that this graph reproduces the structure of Gowalla network satisfactorily. The last diagram on Fig. 5 demonstrates that VDD $\{Q_k\}$ of the calibrated graph describes the network nodes degree distribution adequately.

The quality of the calibration result obtained manually indicates the possibility of its significant refinement by formulating and solving the resulting extreme problem in an appropriate way. There are no «artificial» technical adjustments (complex linear weights and «optimized» sets $\{r_k\}$) and no real mechanisms of the networks formation unexplained logically, which allows us to suppose that the simulated network Gowalla does consist of the sub-networks with an autonomous structure.

It should also be mentioned that the supposition on the success correlation at the edges distribution in the AER component of the calibrated graph seems natural since it correlates with the methods applied in Gowalla network to stimulate the users' activity [8, 9].

## VII. CONCLUSION

The research findings demonstrate that the structure of some social networks is non-uniform. Social network Brighkite can be represented as a composition (combination) of two large components with autonomous structures. Both components are properly described by the preferential attachment graphs. Geosocial networkGowalla can be treated as the composition of an autocorrelated random Erdős–Rényi network and a network described by the preferential attachment graph. The appropriate modification of the Erdős–Rényi graph, an autocorrelated ER graph introduced in the present work, takes into account anautocorrelatedbehaviour of the network users making connections, the behavior affected by the memory of previous successes or failures.

In fact, well-calibrated network graph models and computer experiments with these models would help developers (owners) of the networks to predict their development correctly and to choose effective strategies for controlling network projects. The story [9] of Gowalla network creation, its unexpected loss to competing networks and the subsequent forced sale to Facebook owners, demonstrates the significance of such networks simulation.

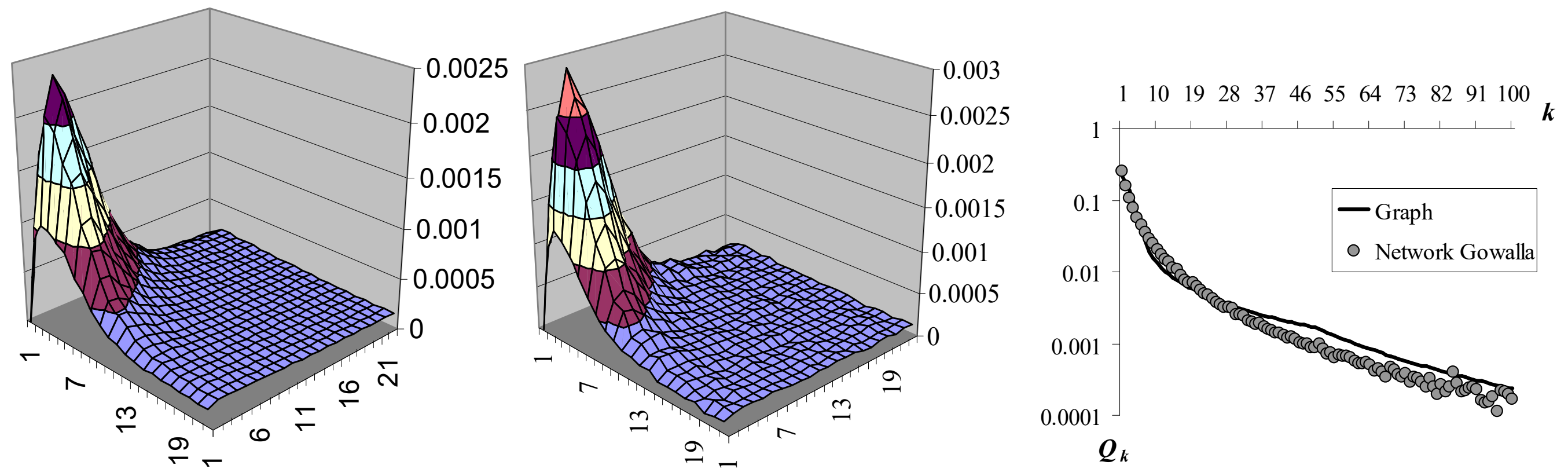

Fig. 5. The calculated edge degree distribution of the calibrated two-component graph is on the left, the edge degree distribution of the calibrated graph obtained through SM is in the center, the graph vertices and network nodes degree distributions are compared on the right